\documentclass{article}
\usepackage[a4paper]{geometry}
\usepackage{amsmath}

\usepackage[sort&compress,numbers]{natbib}
\DeclareMathOperator{\E}{e}
\newcommand{\I}{\mathrm{i}}

\newcommand{\vect}[1]{\boldsymbol{#1}}
\newcommand{\vers}[1]{\hat{\boldsymbol{#1}}}
\newcommand{\tens}[1]{\hat{#1}}
\newcommand{\mat}[1]{\boldsymbol{\mathsf{#1}}}

\newcommand{\tsub}[1]{_{\text{#1}}}
\newcommand{\tsup}[1]{^{\text{#1}}}

\newcommand{\pderiv}[3][]{\frac{\partial^{#1} #2}{\partial #3^{#1}}}

\makeatletter
\newenvironment{rcases}{%
   \matrix@check\rcases\env@rcases
}{%
   \endarray\right\rbrace%
}
\def\env@rcases{%
   \let\@ifnextchar\new@ifnextchar
\left.
   \def\arraystretch{1.2}%
   \array{@{}l@{}l@{}}%
}\makeatother 

\author{Wojciech \'Smigaj, Boris Gralak\thanks{Institut Fresnel, UMR CNRS 6133, 
    13397 Marseille Cedex 20, France.}}
\title{Multiple-scattering method applied to gyrotropic media}
\date{March 16, 2009}

\begin{document}
\maketitle

\noindent In the following, we present the modifications of the multiple-scattering method~\cite{FelbacqJOSAA94,TayebJOSAA97,TayebBook04} necessary for handling two-dimensional systems of scatterers characterised by tensor (relative) permittivity~$\tens\epsilon$ of the form
\begin{equation}
  \label{eq:epsilon}
  \tens\epsilon = 
  \begin{bmatrix}
    \epsilon\tsub t & \I g & 0\\
    -\I g & \epsilon\tsub t & 0\\
    0 & 0 & \epsilon_z
  \end{bmatrix}.
\end{equation}
For simplicity, we assume all media to be nonmagnetic ($\mu \equiv 1$).
We assume that the direction of invariance coincides with the $z$~axis.

With the permittivity given by eq.~(\ref{eq:epsilon}), the Maxwell's curl equations
\begin{subequations}
  \label{eq:maxwell-general}
  \begin{align}
    \vect \nabla \times \vect E &= \I \omega \mu_0 \vect H,\\
    \vect \nabla \times \vect H &= -\I \omega \tens \epsilon \epsilon_0 \vect E
  \end{align}
\end{subequations}
decouple into the equations corresponding to the $E$ polarization ($\vect E \parallel \vers z$), 
\begin{subequations}
  \label{eq:maxwell-E}
  \begin{align}
    \pderiv{E_z}{y} &= \I\omega\mu_0 H_x,\\
    -\pderiv{E_z}{x} &= \I\omega\mu_0 H_y,\\
    \pderiv{H_y}{x} - \pderiv{H_x}{y} &= -\I\omega \epsilon_0 \epsilon_z E_z,
  \end{align}
\end{subequations}
and the $H$ polarization ($\vect H \parallel \vers z$),
\begin{subequations}
  \label{eq:maxwell-H}
  \begin{align}
    \pderiv{E_y}{x} - \pderiv{E_x}{y} &= \I\omega \mu_0 H_z,\\
    \pderiv{H_z}{y} &= -\I\omega\epsilon_0(\epsilon\tsub t E_x + \I g E_y),\\
    -\pderiv{H_z}{x} &= -\I\omega\epsilon_0(-\I g E_x + \epsilon\tsub t E_y).
  \end{align}
\end{subequations}
In the following, we restrict our attention to the case of the $H$ polarization, where the effects of anisotropy are present. From eqs.~(\ref{eq:maxwell-H}) we obtain
\begin{subequations}
  \label{eq:E-components}
  \begin{align}
    E_x &= \frac{1}{\I\omega\epsilon_0(g^2 - \epsilon\tsub t^2)}
    \biggl(\I g \pderiv{H_z}{x} + \epsilon\tsub t \pderiv{H_z}{y}\biggr),\\
    E_y &= \frac{1}{\I\omega\epsilon_0(g^2 - \epsilon\tsub t^2)}
    \biggl(-\epsilon\tsub t \pderiv{H_z}{x} + \I g \pderiv{H_z}{y}\biggr)
  \end{align}
\end{subequations}
and
\begin{equation}
  \label{eq:helmholtz}
  \pderiv[2]{H_z}{x} + \pderiv[2]{H_z}{y}+ 
  \epsilon_0 \mu_0 \biggl(\epsilon\tsub t - \frac{g^2}{\epsilon\tsub t}\biggr)
  \omega^2 H_z = 0.
\end{equation}
Evidently, the $H_z$ field fulfils the Helmholtz equation $\nabla^2H_z + k^2 H_z = 0$ with $k^2 =  \epsilon_0 \mu_0 (\epsilon\tsub t - g^2/\epsilon\tsub t)$. This is the basic assumption of the multiple-scattering formalism, which we can therefore apply to the analysis of the considered system. With respect to the multiple-scattering method as described in Ref.~\cite{TayebBook04}, we need to make only two changes: (1) replacement of the formula $n = \sqrt{\epsilon\tsub t\vphantom{I}}$ for the refractive index of a medium by $n = \sqrt{\epsilon\tsub t - g^2/\epsilon\tsub t}$, (2) modification of the formula for the scattering matrix of a cylinder. This formula can be derived easily by imposing the conditions of continuity of the tangential components of the electric and magnetic fields on the surface of a cylinder, which we do in the following. 

For the case of $H$ polarization, the tangential components in question are $H_z$ and $E_\phi$, so we need to impose that
\begin{equation}
  \label{eq:bnd-conds}
  \begin{rcases}
    H_z\tsup{int}(R, \phi) = H_z\tsup{ext}(R, \phi)\\[\jot]
    E_\phi\tsup{int}(R, \phi) = E_\phi\tsup{ext}(R, \phi)
  \end{rcases}
  \qquad\text{for all $\phi\in[0,2\pi)$,}
\end{equation}
where $R$ denotes the radius of the cylinder and the superscripts ``int'' and ``ext'' refer to field expansions in the cylinder interior and exterior, respectively.  Equations~(\ref{eq:E-components}) imply that
\begin{equation}
  \label{eq:E-phi}
  E_\phi = \frac{1}{\I\omega \epsilon_0(g^2 - \epsilon\tsub t^2)}
  \biggl(-\epsilon\tsub t \frac{\partial H_z}{\partial r} +
  \frac {\I g} r \frac{\partial H_z}{\partial\phi}\biggr).
\end{equation}
From the multiple-scattering theory we know that the magnetic field~$H_z$ inside and outside the cylinder is given by appropriate Fourier-Bessel expansions,
\begin{subequations}
  \label{eq:H-z-expansions}
  \begin{align}
    H_z\tsup{int}(r, \phi) &= 
    \sum_m c_m \underbrace{J_m(k\tsub{int}r)}_{\equiv C^H_m} \E^{\I m\phi},\\
    H_z\tsup{ext}(r, \phi) &= 
    \sum_m \bigl[a_m \underbrace{J_m(k\tsub{ext}r)}_{\equiv A^H_m} {}+
    b_m \underbrace{H_m^{(1)}(k\tsub{ext}r)}_{\equiv B^H_m} \bigr]
    \E^{\I m\phi},
  \end{align}
\end{subequations}
hence
\begin{subequations}
  \label{eq:E-phi-expansions}
  \begin{align}
    E_\phi\tsup{int}(R, \phi) &= 
    -\frac{1}{\I\omega\epsilon_0}
    \sum_m c_m 
    \underbrace{\frac{(m g\tsub{int}/R) J_m(k\tsub{int}R) + 
        \epsilon\tsub{t,int}k\tsub{int} J_m'(k\tsub{int}R)}
      {g\tsub{int}^2-\epsilon\tsub{t,int}^2}}_{\equiv C^E_m}
    \E^{\I m\phi},\\\nonumber
    E_\phi\tsup{ext}(R, \phi) &= 
    -\frac{1}{\I\omega\epsilon_0}
    \biggl\{\sum_m a_m 
    \underbrace{\frac{(m g\tsub{ext}/R)J_m(k\tsub{ext}R) + 
      \epsilon\tsub{t,ext} k\tsub{ext} J_m'(k\tsub{ext}R)}
    {g\tsub{ext}^2-\epsilon\tsub{t,ext}^2}}_{\equiv A^E_m}
    \\ &\qquad\quad\qquad+
    \sum_m b_m 
    \underbrace{\frac{(m g\tsub{ext}/R)H^{(1)}_m(k\tsub{ext}R) + 
        \epsilon\tsub{t,ext} k\tsub{ext} H^{(1)\prime}_m(k\tsub{ext}R)}
      {g\tsub{ext}^2-\epsilon\tsub{t,ext}^2}}_{\equiv B^E_m}
    \biggr\}\E^{\I m\phi}.
  \end{align}
\end{subequations}
Substituting these formulas to eqs.~(\ref{eq:bnd-conds}) and solving for $b_m$ and $c_m$, we obtain
\begin{subequations}
  \begin{align}
    b_m &= -\frac{A^H_m C^E_m - A^E_m C^H_m}{B^H_m C^E_m - B^E_m C^H_m} a_m,\\
    c_m &= -\frac{A^H_m B^E_m - A^E_m B^H_m}{B^H_m C^E_m - B^E_m C^H_m} a_m.
  \end{align}
\end{subequations}
We conclude that the scattering matrix~$\mat S$ of a cylinder of circular cross-section is diagonal and its elements are given by
\begin{equation}
  \label{eq:S-matrix}
  S_{ij} =
  -\frac{A^H_m C^E_m - A^E_m C^H_m}{B^H_m C^E_m - B^E_m C^H_m} \delta_{ij},
\end{equation}
where $\delta_{ij}$ is the Kronecker delta and the symbols $A^E_m$ etc.\ are defined in eqs.\ (\ref{eq:H-z-expansions}) and (\ref{eq:E-phi-expansions}).

\end{document}